\documentclass[twocolumn,showpacs,preprintnumbers,amsmath,amssymb,prb,superscriptaddress]{revtex4}
\usepackage{graphicx}
\usepackage{epsf}
\usepackage{bm}
\usepackage{hyperref}
\usepackage{color}
\begin{document}

\title{Contactless Measurement of AC Conductance in Quantum Hall
Structures}

\author{I.~L.~Drichko}
\affiliation{A.~F.~Ioffe Physical-Technical Institute of the Russian
Academy of Sciences,  194021 St. Petersburg, Russia}
\author{A. M. Diakonov}
\affiliation{A.~F.~Ioffe Physical-Technical Institute of the Russian
Academy of Sciences,  194021 St. Petersburg, Russia}
\author{V.~A.~Malysh}
\affiliation{A.~F.~Ioffe Physical-Technical Institute of the Russian
Academy of Sciences,  194021 St. Petersburg, Russia}
\author{I.~Yu.~Smirnov}
\affiliation{A.~F.~Ioffe Physical-Technical Institute of the Russian
Academy of Sciences,  194021 St. Petersburg, Russia}
\author{Y.~M.~Galperin}
\affiliation{Department of Physics, University of Oslo, PO Box 1048
Blindern, 0316 Oslo, Norway} \affiliation{A.~F.~Ioffe
Physical-Technical Institute of the Russian Academy of Sciences,
194021 St. Petersburg, Russia}
\author{N.~D.~Ilyinskaya}
\affiliation{A.~F.~Ioffe Physical-Technical Institute of the Russian
Academy of Sciences,  194021 St. Petersburg, Russia}
\author{A.~A.~Usikova}
\affiliation{A.~F.~Ioffe Physical-Technical Institute of the Russian
Academy of Sciences,  194021 St. Petersburg, Russia}
\author{M.~Kummer}
\affiliation{Laboratorium f$\ddot{u}$r Festk\"{o}rperphysik ETH
Z\"{u}rich, CH-8093 Z\"{u}rich Switzerland}
\author{H.~von~K{\"a}nel}
\affiliation{Laboratorium f$\ddot{u}$r Festk\"{o}rperphysik ETH
Z\"{u}rich, CH-8093 Z\"{u}rich Switzerland}
\begin{abstract}
We report a procedure to determine the frequency-dependent
conductance of quantum Hall structures in a broad frequency domain.
The procedure is based on the combination of two known probeless
methods -- acoustic spectroscopy and microwave spectroscopy. By
using the acoustic spectroscopy,
 we study the low-frequency attenuation and phase shift of a
surface acoustic wave in a piezoelectric crystal in the vicinity of
the electron (hole) layer. The electronic contribution is resolved
using its dependence on a transverse magnetic field.  At high
frequencies, we study the attenuation of an electromagnetic wave in
a coplanar waveguide. To quantitatively calibrate these data, we use
the fact that in the quantum-Hall-effect regime the conductance at
the maxima of its magnetic field dependence is determined by
extended states. Therefore, it should be frequency independent in a
broad frequency domain.  The procedure is verified by studies of a
well-characterized  $p$-SiGe/Ge/SiGe heterostructure.
\end{abstract}

\maketitle

\section{Introduction}
The dynamics of charge carriers in low-dimensional quantum structures has
been in the focus of interest for many years.  A special role in its study is played by
probeless methods allowing the influence of contacts to be avoided.

In one of these methods the attenuation and the phase shift of
a surface acoustic
 wave (SAW) are measured. These are caused by charge carriers in a two-dimensional (2D) layer
 located close to the surface supporting the SAW.  To the best of our knowledge, the first results
 of acoustic studies of 2D electron systems were reported in Ref.~\onlinecite{Wixforth1986}. In that paper,
 the interaction of a SAW with the 2D electrons in a GaAs/AlGaAs heterostructure was investigated in the integer
 quantum Hall effect (IQHE) regime. In this work, a SAW was
 excited and propagated directly on the surface of a piezoelectric GaAs/AlGaAs
 sample. This procedure caused the sample to be somewhat mechanically stressed and thereby deformed.
Later the authors suggested a configuration in which no mechanical
deformation is transferred
 from the piezoelectric substrate to the sample, such that
 only the electrical field matters.~\cite{Wixforth1989}
This allowed to determine the complex  AC conductance, $\sigma
(\omega)$, from the measured SAW attenuation and phase shift.  A
relevant theoretical model developed in
Refs.~\onlinecite{Efros1990,Simon1996,Kagan1997} allowed extracting
quantitative information, as was shown in subsequent
works.~\cite{Drichko2000,Drichko2005,Drichko2008,Drichko2009,Drichko2013}
However, the acoustic method has an upper frequency limit,associated
with technological problems in producing the inter-digital
transducers (IDTs) used for the excitation and detection of the SAW,
see Sec.~\ref{experiment_b}.

At the same time, measurements of the electron response in
low-dimensional systems to high-frequency perturbations are
especially important, as being related to several intrinsic
properties of low-dimensional systems. Among these properties are
peculiarities of electron localization, collective modes and their
pinning, mechanisms of integer and fractional quantum Hall effects,
etc.  A powerful approach is provided by microwave spectroscopy
(MWS) suggested in Ref.~\onlinecite{Engel1993}.  Various
modifications of this method, including probeless ones, were
developed in
Refs.~\onlinecite{Stone2012,Endo2013,Chen_thesis,Stone_thesis}. The
MWS provides a much broader frequency domain compared with acoustic
spectroscopy~(AS) - from hundreds of megahertz to dozens of
gigahertz. It is very efficient for studying the dependence of the
electron response on magnetic field, temperature, etc.  However, it
is rather difficult to calibrate the measured AC conductance in
absolute units, which is why the results of many papers on this
subject are presented in relative units.

The aim of the present work is to compare the results of  microwave
measurements of $\sigma (\omega)$ with those obtained using AS,
as well as DC transport measurements. This comparison will
facilitate the calibration of the dynamical
electromagnetic response of the low-dimensional electron gas in
absolute units.

The paper is organized as follows. In Sec.~\ref{experiment} we
describe the sample, as well as both AS
 and MWS.  The results are reported in Sec.~\ref{results} and discussed
 in Sec.~\ref{discussion}.
\section{Experiment} \label{experiment}
\subsection{Sample}\label{sample}
\begin{figure}[h!]
\centerline{
\includegraphics[width=.4\columnwidth]{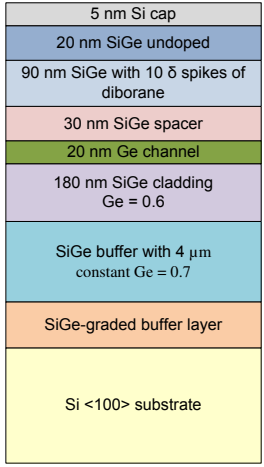}
}
\caption{Structure of the sample. \label{fig1}}
\end{figure}
Since it is very important to use well-characterized samples we have
chosen the \textit{p}-SiGe/Ge/SiGe heterostructure  (K6016)
investigated earlier in Refs.~\onlinecite{Drichko2009,Drichko2013}
by means of AS. The structure of the sample is schematically shown
in Fig.~\ref{fig1}. The sample was made by low-energy
plasma-enhanced chemical vapor deposition (LEPECVD),
see~\cite{lepecvd} for details. The active part of the sample is a
2D hole channel formed in a strained Ge layer. The density and
mobility of holes are $p=6\times10^{11}$~cm$^{-2}$ and
$\mu_p=6\times 10^4$~cm$^2$/(V$\cdot$ s), respectively, at 4.2 K.

\subsection{Experimental methods} \label{methods}
In the following, we briefly explain the two methods employed for the AC conductance measurements - one
(AS) based on the propagation of the SAWs and the other on microwave spectroscopy
(MWS).

\subsubsection{Acoustic spectroscopy} \label{experiment_b}

\begin{figure}[b]
\centerline{
\includegraphics[width=.8\columnwidth]{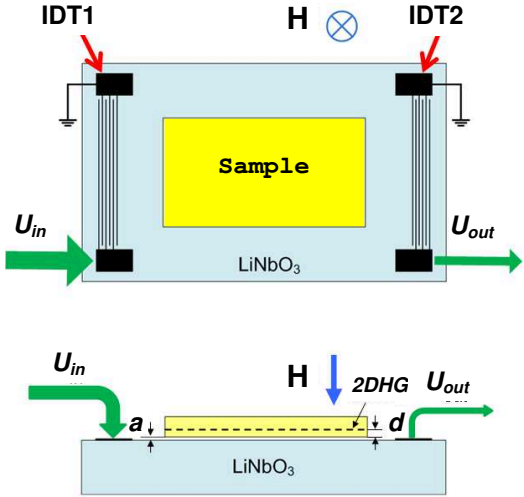}
}
\caption{Sketch of the AS setup. \label{fig2}}
\end{figure}

In Fig.~\ref{fig2} is shown a sketch of the acoustic method.
A SAW is excited on the surface of a piezoelectric LiNbO$_3$ crystal
by electromagnetic pulses applied to an interdigital transducer IDT1
and detected by another interdigital transducer IDT2 placed on the
same surface. The SAW, generated by the piezoelectric effect in
LiNbO$_3$, is accompanied by a travelling electromagnetic wave. The
sample is pressed onto the surface by a spring. As a result, the
electric field of the travelling wave penetrates into the 2D hole
gas and interacts with holes.  This interaction leads to an
attenuation, $\Gamma$,  and a phase shift of the SAW, both of which
are measured.  The latter manifests itself as a renormalization of
the SAW phase velocity, $v$. It is important to note that both real
and imaginary parts of the conductance vanish in a strong
perpendicular magnetic field. This creates the possibility to single
out the electron contribution by subtracting the zero-field values
of $\Gamma$ and $\Delta v$ from those in a strong magnetic field.
The expressions for these differences, as well as the parameters
which we use for finding the conductances,  are given in
Appendix~\ref{A} as Eqs.~\eqref{eq:01} and \eqref{eq:02}. They are
based on the derivation given in Ref.~\onlinecite{Kagan1997}.

In principle, Equations~\eqref{eq:01} and \eqref{eq:02} allow
finding both the real and imaginary parts of the complex conductance
from the measured $\Gamma$ and $\Delta v/v$.

\subsubsection{Microwave spectroscopy} \label{microwave}

\begin{figure}[t]
\centering
\includegraphics[width=0.8\columnwidth]{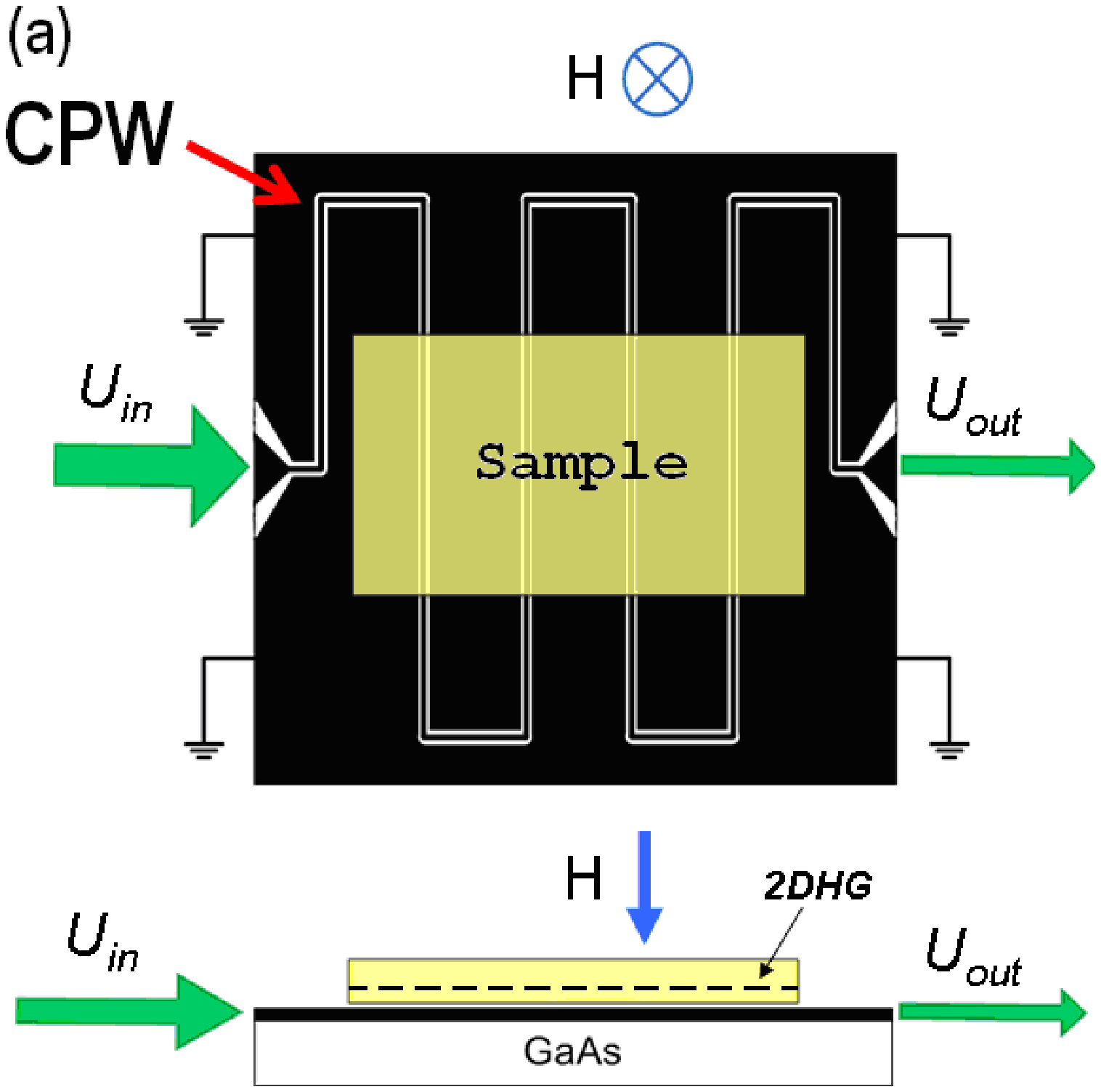} \\[1mm]
\includegraphics[width=.8\columnwidth]{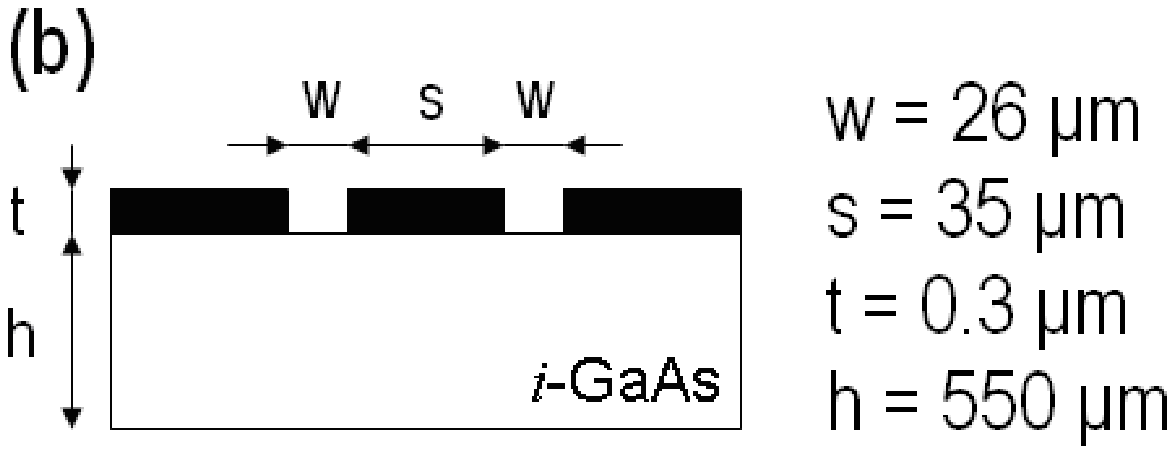} \\[1mm]
\includegraphics[width=0.6\columnwidth]{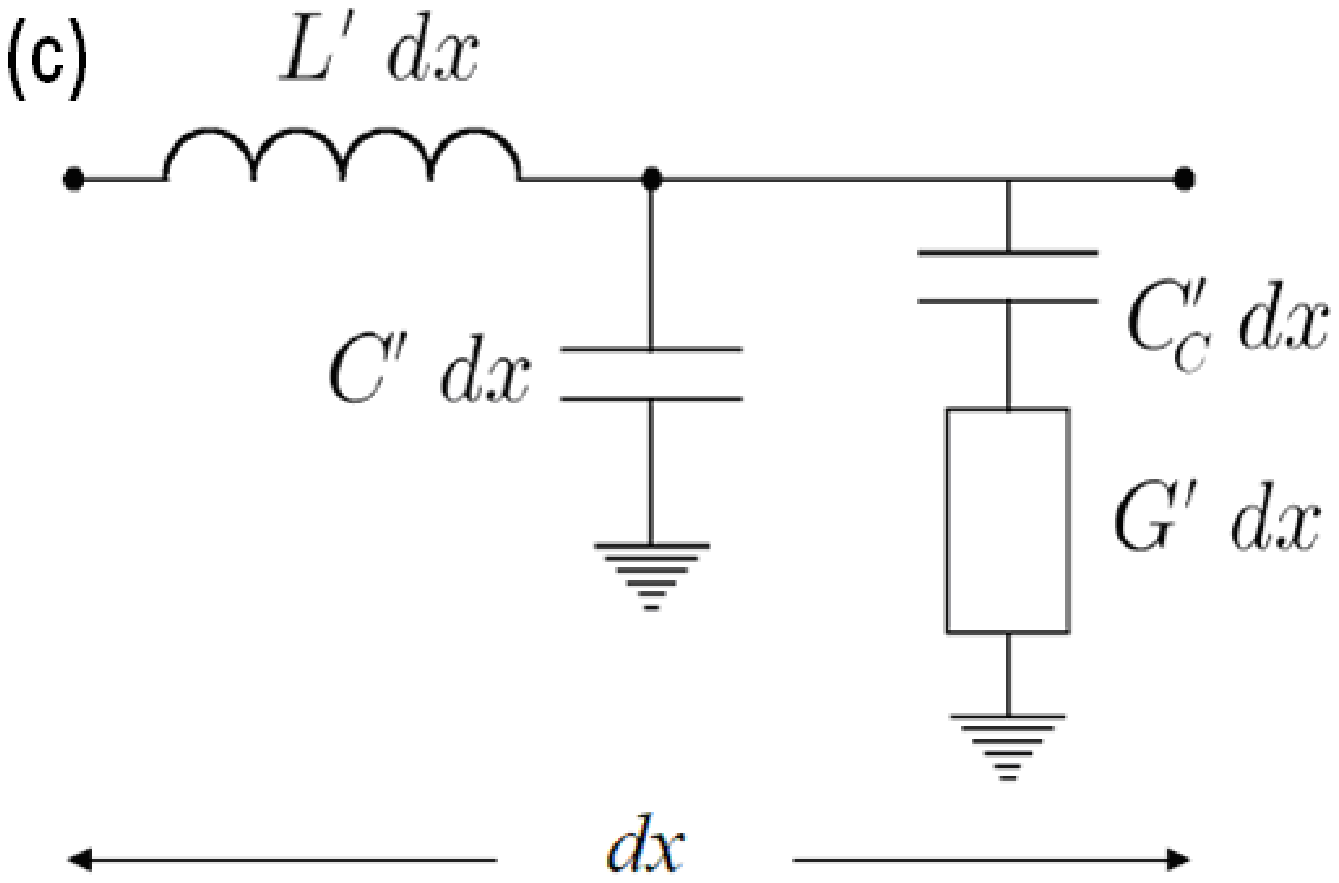}
\caption{(a) - Sketch of the MWS setup; (b) - Sizes
of the elements of the CPW. (c) - A simple transmission line model for
 using the CPW to measure the AC conductance of the 2D hole layer.
 \label{fig3}}
\end{figure}

In Fig.~\ref{fig3} is shown a sketch of the experimental setup for
microwave spectroscopy. In this case, the sample is placed on a
meandered coplanar waveguide (CPW) formed on the surface of an
insulating GaAs substrate.  The microwave pulses applied to the CPW
center conductor excite a quasi-transverse electromagnetic mode
(quasi-TEM mode). Similarly to the case of a SAW, the interaction
with holes in the 2D layer leads to the attenuation of this mode, as
well as to a change of its phase. Again, both effects are due to the
complex conductance of the 2D layer. To separate these
contributions, we will subtract the results in a transverse magnetic
field from the zero-field ones.

To relate the measurable quantities with the complex conductance,  a
simple transmission line model is used as outlined in Fig.~\ref{fig3}c. Here
$L^\prime$  is the inductance of the center conductor per unit
length and $C^\prime$ the capacitance between the center conductor and
ground (side plane) per unit length. The 2D layer constitutes
a shunt admittance from the CPW center conductor to the ground.
$C^\prime =sC_g$  is the capacitance per unit length from the center conductor (of
width $s$) to the 2D layer (located at a distance $d$ below the
surface), where $C_g=\varepsilon/d$ is the
capacitance per unit area. If $\xi =\sqrt{|\sigma_{\alpha
\alpha}|/(\omega C_g)} =\sqrt{|\sigma_{\alpha \alpha}|d/\varepsilon}
\ll w$, the microwave electric field is mainly confined in the
slots, and the shunt capacitance ($C^\prime_c$ term) has a
negligible contribution compared to that of the shunt conductance
($G^\prime$ term, where $G^\prime =2\sigma_{\alpha \alpha}/w$ is the
conductance per unit length of the 2D layer under both slots (thus
the factor of 2)). In this case, the CPW simply acts as contacts to the
2D hole layer. Here $\sigma_{\alpha \alpha}$ is the conductance in
the direction of the electric field.

The wave attenuation is given as~\cite{Simons,Chen_thesis}:
\begin{equation}\label{CPW-r}
\Gamma= - \frac{1}{2l}\ln  \left(\frac{P_{\text{out}}}{P_{\text{in}}}\right) =
\Re \left[\sqrt{ i \omega L^\prime (i\omega C^\prime +G^\prime)}\right]\,.
\end{equation}
Substituting the expressions for $L^\prime$, $C^\prime$ and
$G^\prime$  (given, e.g., by Eqs. (D.1)-(D.3) in Appendix D
of~\cite{Chen_thesis},), we can express $\Re \sigma_{\alpha \alpha}$
through the attenuation coefficient,  $\Gamma$. This expression can
be cast in the form (cf. with~\cite{Endo2013}):
\begin{equation} \label{sigma_full}
\frac{Z_0 \Re(\sigma_{\alpha \alpha})}{w}=\Gamma \sqrt{1+
\left(\frac{v_{\text{ph}}\Gamma}{\omega}\right)^2}\, .
\end{equation}
Here $v_{\text{ph}}=1/\sqrt{L^\prime C^\prime}
=c\sqrt{2/(1+\varepsilon_1)}$ is the phase velocity of the wave, $c$
- is the vacuum velocity of light, $\varepsilon_1$ is the dielectric
constant of $i$-GaAs, and $Z_0 =\sqrt{L^\prime/C^\prime}$ is the
characteristic impedance. In our experiment  the parameters of the
CPW are selected in order to have  $v_{\text{ph}}=1.02 \times
10^8$~m/s and $Z_0 = 50$~Ohm, see Fig.~\ref{fig3}.

Equation~\eqref{sigma_full} is valid under the following
conditions~\cite{Engel1993}: (i) There are no reflections at the CPW
edges; (ii) The AC current is concentrated inside the slots, i.~e.,~the
width of the slot, $w$, should exceed the wave penetration depth in
the stripes. The inequality can be written as
\begin{equation} \label{concentration}
w \gtrsim \sqrt{\frac{\sigma_1}{\pi f C_c}}\, , \quad C_c=
\varkappa_0\frac{\varepsilon_s\varepsilon_0}{a \varepsilon_s+d\varepsilon_0}\, .
\end{equation}
Here $C_c$ is the capacitance between the metallic grounded stripe
and the 2D layer per unit area, $\varkappa_0= 8.854\times
10^{-12}$~F/m is the vacuum permittivity,  $a$ is the clearance
between the sample and the $i$-GaAs surface; $\varepsilon_0$=1 and
$\varepsilon_s$=16.2 are the dielectric  constants, of vacuum, and
of the semiconductor, respectively. Putting $a= 1$~$\mu$m and
$d=0.145$~$\mu$m, we get $C_c=0.88\times 10^{-5}$~F/m$^2$. Then the
inequality~\eqref{concentration} can be rewritten as
\begin{equation}\label{c1}
\sigma_1 \ (\text {Ohm}^{-1}) \lesssim 1.87\times 10^{-8} f  \ (\text{MHz})\, ,
\end{equation}
where we have substituted $w=26$~$\mu$m. The inequality~\eqref{c1}
is a serious limitation for the quantitative determination of
$\sigma_1$ for samples with relatively large conductance.
Therefore, one should be careful while extracting quantitative
information from electromagnetic measurements, especially at
relatively low frequencies.  In what follows we will report our
procedure for extracting $\sigma_1 (\omega)$ in a broad frequency
domain by combining the results of acoustic and electromagnetic
measurements.

\section{Results} \label{results}

\subsubsection{Acoustic spectroscopy}
\begin{figure}[t]
\centering
\includegraphics[width=.8\columnwidth]{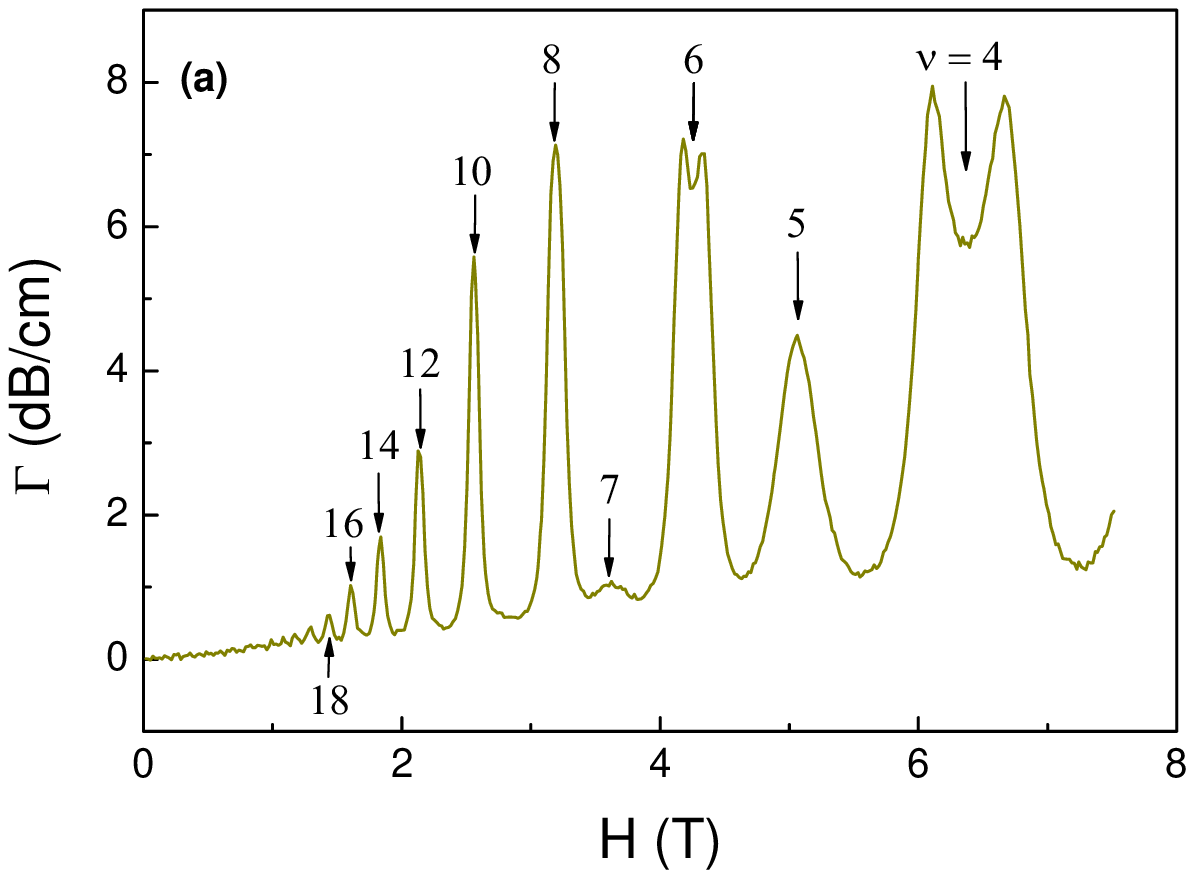} \\
\includegraphics[width=.8\columnwidth]{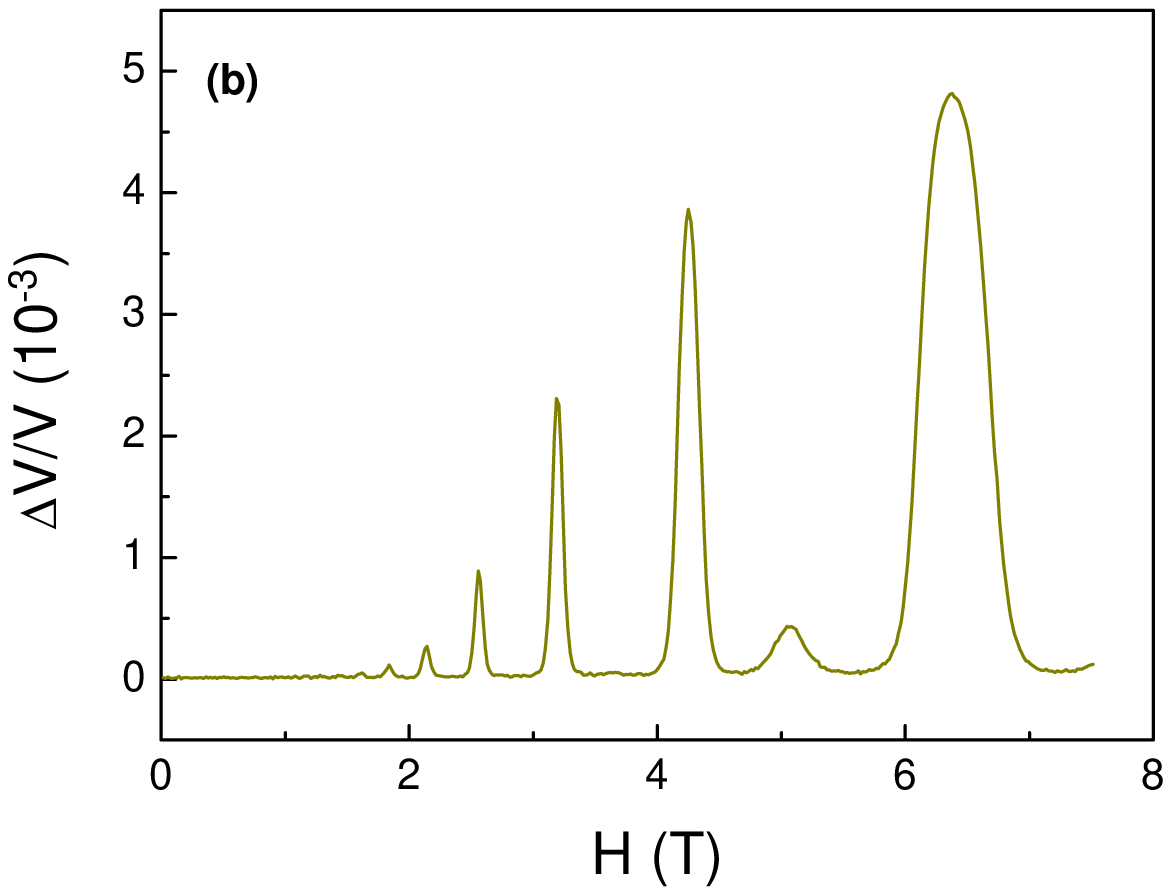}\\
\includegraphics[width=.8\columnwidth]{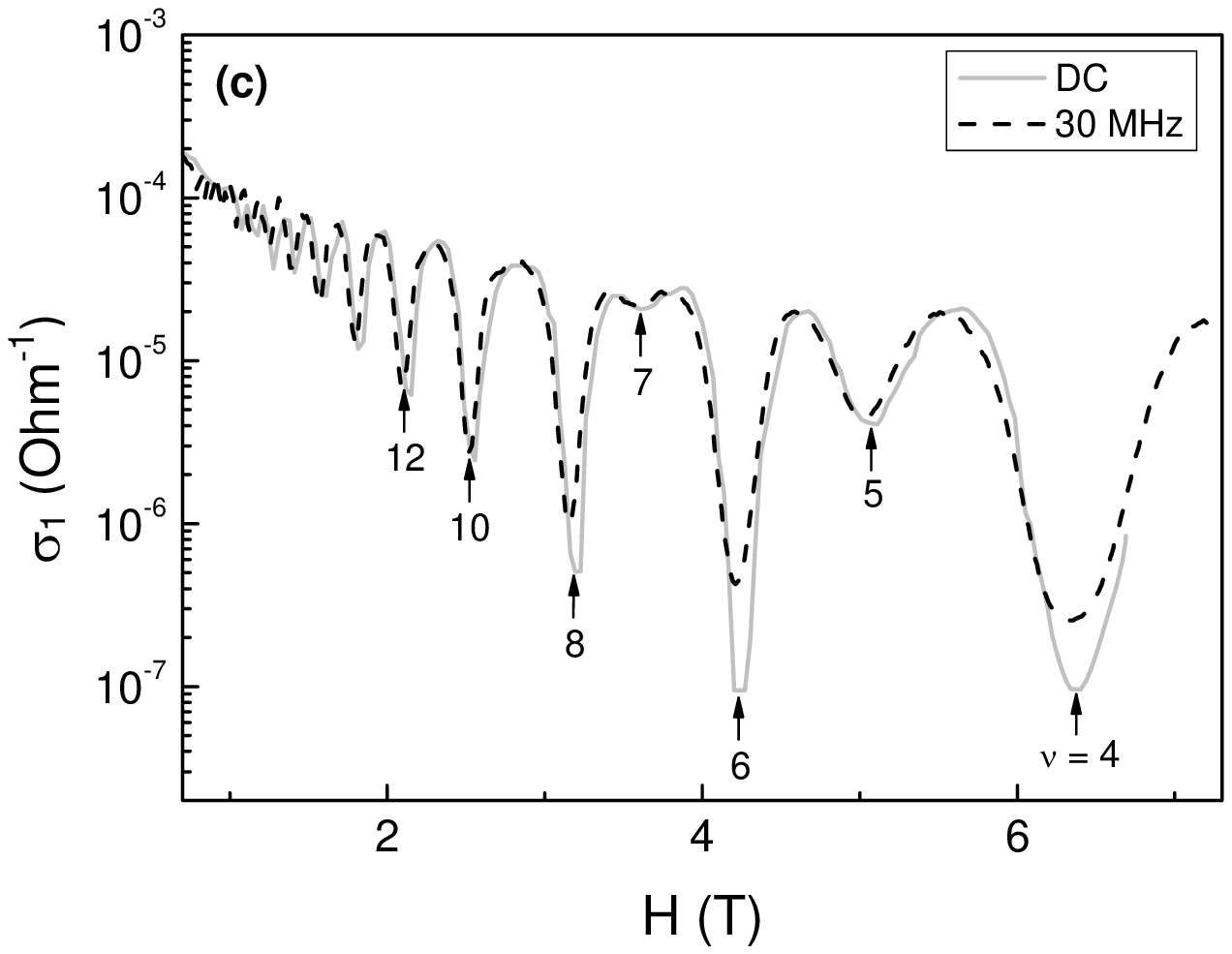}
\caption{Magnetic field dependences of the acoustic absorption
$\Gamma$ (a), the variation of
the velocity $\Delta v/v$ (b) and the conductance $\sigma_1$,
 as obtained from these experimental data (c).
$f=30$~MHz, $T=1.7$~K. Values of the filling factor $\nu$ are shown close to corresponding conductance
minima. The gray line in panel (c) shows the DC conductance, $\sigma_0(H)$.  \label{fig4}}
\end{figure}

Acoustic spectroscopy is most suitable for low frequencies, the
upper frequency limit being mainly defined by design of the IDT. Shown
in Fig.~\ref{fig4} are magnetic field dependences of the
attenuation, $\Gamma$ (a), and the SAW velocity, $\Delta v/v$ (b) at
a frequency $f=\omega/2\pi=30$~MHz and temperature $T=1.7$~K. The
filling factors are shown by arrows.
Using the procedure based on the expressions given in
Appendix~\ref{A} and described in detail
in~\cite{Drichko2000,Drichko2008,Drichko2013}, we can extract both
the real and imaginary parts of the complex AC conductance.  The
extracted magnetic field dependence of $\sigma_1$ is shown in
Fig.~\ref{fig4}c.

\subsubsection{Microwave spectroscopy} \label{em}

\begin{figure}[h!]
\centering
\includegraphics[width=.9\columnwidth]{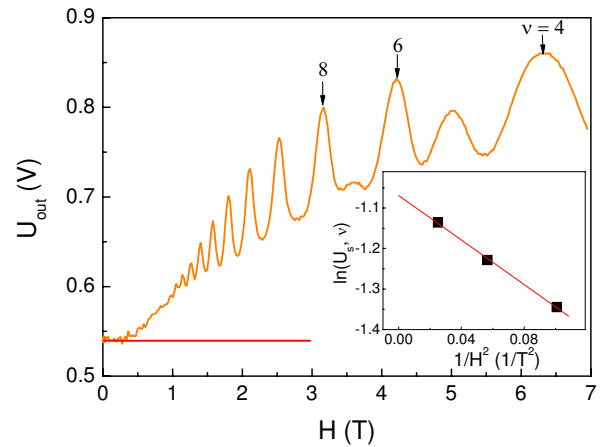}
\caption{ Magnetic field dependence of  $U_{\text{out}}$.
Inset - Dependence of $\ln (U_s, V)$ on $H^{-2}$.
 $f = 1102$~MHz, $T=1.7$~K.  \label{fig7}}
\end{figure}

Shown in Fig.~\ref{fig7}a is the magnetic field dependence of the
output signal, $U_{\text{out}}$ of the CPW at a frequency of 1102~MHz
and a temperature of 1.7~K. The sample is the same as that used for the
acoustic measurements.

To relate the output signal to the sample conductance, one also
needs to know the input signal, $U_{\text{in}}$.  This is not a
trivial task since we observed a significant background signal
independent of the magnetic field.  We attribute this signal to some
leakage in the structure and present the total output signal as
$U_{\text{out}}=U_s(H)+U_l$.  Here $U_s$ is the signal having
interacted with the sample, while $U_l$ (shown by solid red line in
Fig.~\ref{fig7}) represents the leakage. According to our
measurements, the magnetic field dependence of the phase shift of
the total output signal is relatively weak, the field-dependent
shift being 20-50$^\circ$. Therefore, we simply subtract the
background amplitude,~i.e.~$U_l$, from the total output amplitude.

Now we take into account that the oscillations of $U_s$ versus
magnetic field are caused by oscillations of the diagonal
conductance  in the regime of the integer quantum Hall effect.  As is
well known, the maxima of the diagonal conductance correspond to extended
states close to the Landau level centers. At the same time, the wave
frequencies are much below that the typical electron relaxation rate,
$\omega \ll \tau^{-1}$. Therefore, one can expect that the maximal
values of the AC conductance  should coincide with the values of
the static conductance at the same magnetic field. This is illustrated
in Fig.~\ref{fig4}c.

At the same time, the minima of the AC conductance in the IQHE
regime are determined by  hopping between localized states, see,
e.~g.,~Ref.\onlinecite{Polyakov1993} and references therein. The
hopping AC conductance is also suppressed by an external transverse
magnetic field, and in strong magnetic fields $\sigma_1 (\omega)
\propto H^{-2}$ (with logarithmic accuracy), see,
e.g.,~Ref.\onlinecite{Galperin1991-review}.  This dependence is
experimentally confirmed, and we use it to find $U_{\text{in}}$. An
example including 3 signal maxima or conductance minima,
corresponding  to the filling factors $\nu = 4,6,8$ is shown in the
inser of Fig.~\ref{fig7}a (inset), where we have plotted $\ln (U_s,
\text{V})$ versus $H^{-2}$.  The intercept of the straight line fit
with the ordinate axis provides the value of  $\ln (U_{\text{in}},
\text{V})=-1.077$ corresponding to $U_{\text{in}}=0.34$~V. Knowing
$U_{\text{in}}$, we then find the real part of the conductance from
Eq.~\eqref{sigma_full}.

\begin{figure}[h!]
\centering
\includegraphics[width=0.8\columnwidth]{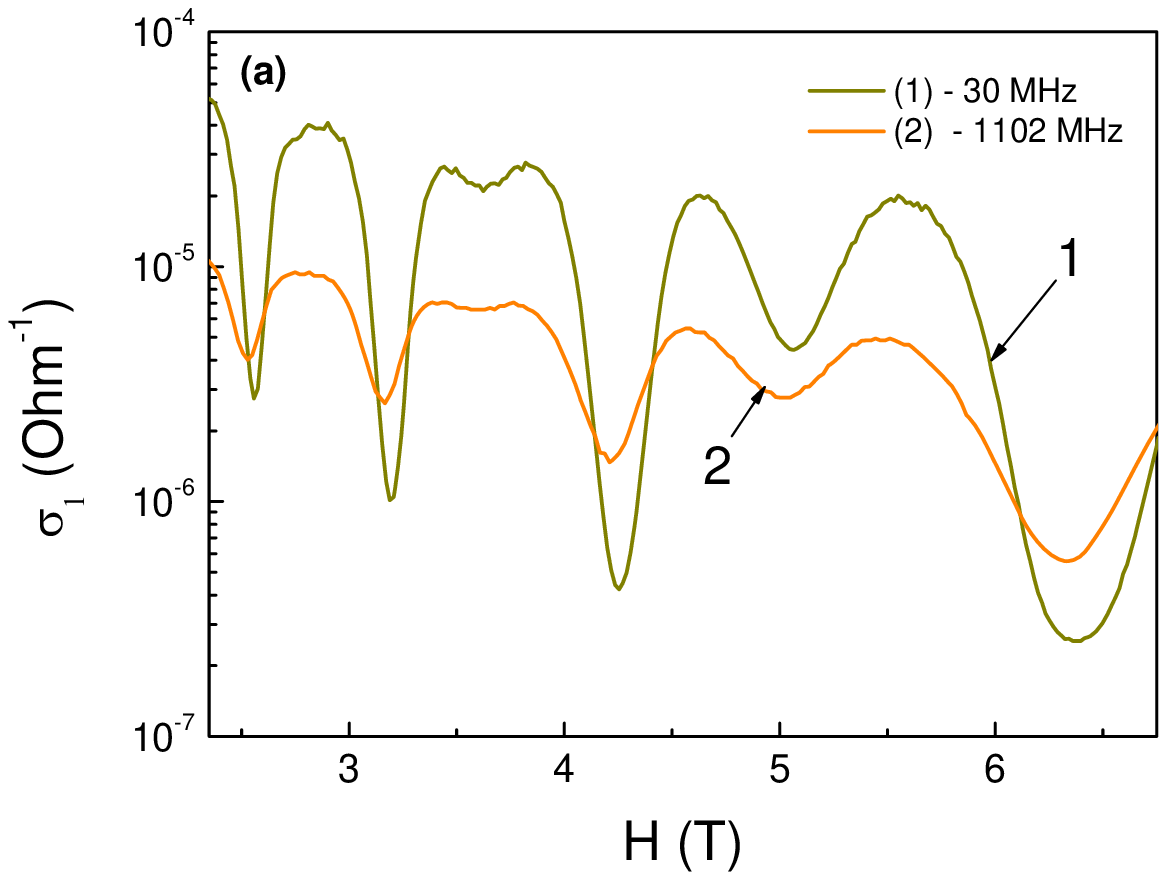} \\
\includegraphics[width=0.8\columnwidth]{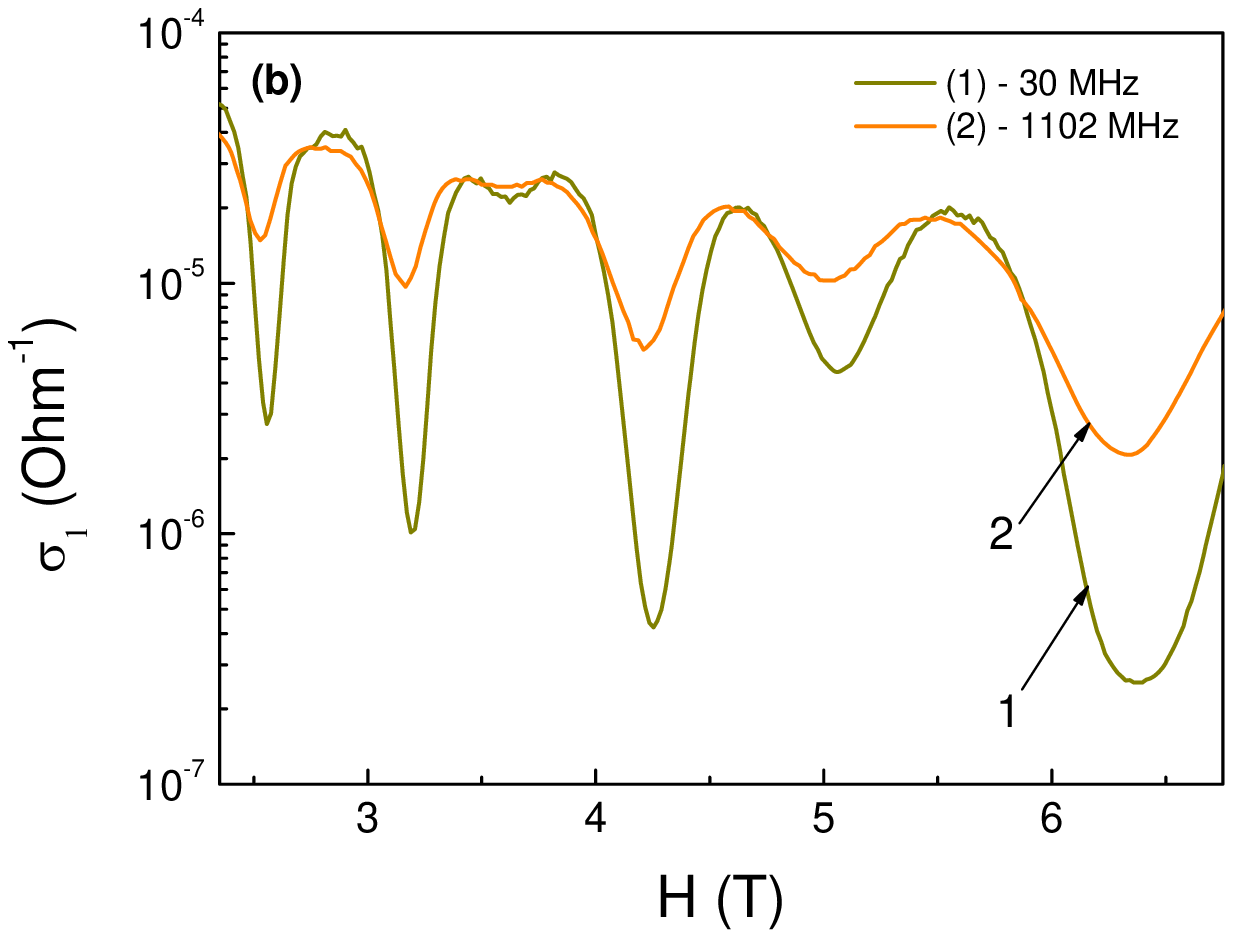} \\[0.05in]
\includegraphics[width=0.8\columnwidth]{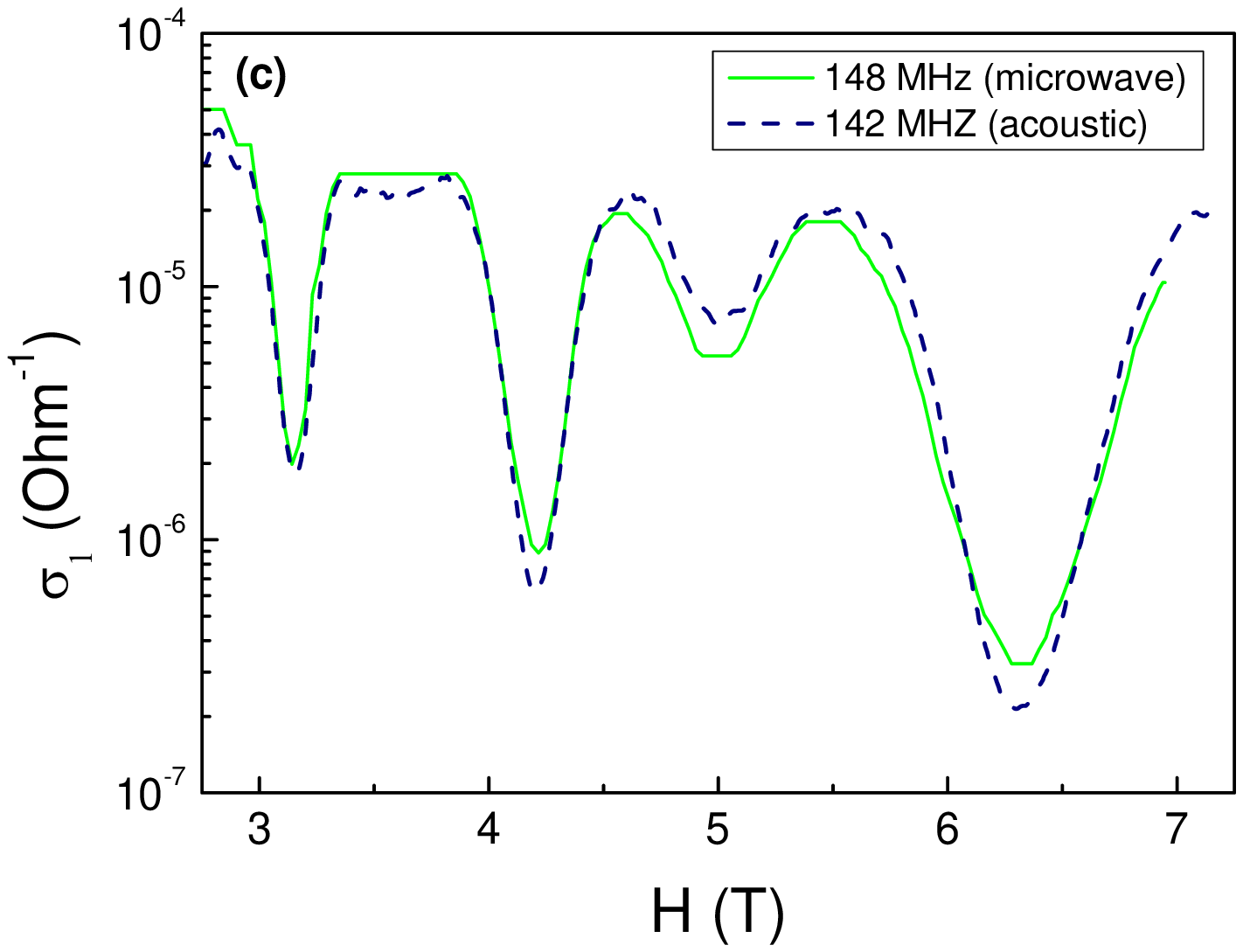}
\caption{(a)~Magnetic field dependences of the conductance $\sigma_1$
at $T=1.7$~K extracted from AS and MWS. The frequencies are 30 and 1102 MHz,
respectively.
(b)~The same dependences, but the results of microwave measurements are multiplied
by the factor $\mathcal{K}=3.7$.
(c)~Results of acoustic and electromagnetic measurements
of frequencies of 142~MHz  and 148~MHz,
respectively.
  \label{fig:fig4_new}}
\end{figure}

 The results for $\sigma_1 (H)$ obtained by AS for $f=30$~MHz and
 MWS for $f=1102$~MHz  are shown in Fig.~\ref{fig:fig4_new}a.

According to our estimates,  the inequality~\eqref{c1} needed for
the validity of Eq.~\eqref{sigma_full} is met only within an order of
magnitude.  Therefore, it is hard to expect high numerical accuracy
of the values of $\sigma_1$ extracted from the CPW measurements.
However, we know that at the maxima of the $\sigma_1 (H)$ curves the
conductance should coincide with the static one for any frequency,
so they should be \textit{frequency-independent}. Therefore, we
suggest rescaling the curve by some factor, $\mathcal{K}(\omega,T)$,
such that the values of $\sigma_1(H)$ at the maxima will be equal to
the static conductance, $\sigma_0 (T)$, for the same temperature.
The results of low-frequency acoustic measurements do not need such
rescaling since, as shown in Fig.~\ref{fig:fig4_new}c, is the required frequency independence of the maxima is fulfilled
automatically.  The result of such rescaling for MWS at $f=1102$~MHz  is
shown in Fig.~\ref{fig:fig4_new}b. The scale factor turns out to
be $\mathcal{K}(1102~\text{MHz}, 1.7~\text{K}) = 3.7$.

At the same time, the values of $\sigma_1$ at the minima are
strongly depend on frequency. This is natural, because a
hopping transport mechanism is valid for magnetic field values
far from the maxima of $\sigma_1(H)$, which inevitably
depends on frequency.  Based on the above considerations,
we use the
following procedure to analyze $\sigma_1 (\omega)$ at the
conductance minima: (i) For each frequency and temperature, after
subtracting the $H$-independent background $U_l$, we determine the
input signal, $U_{\text{in}}$, using the procedure described
earlier; (ii) Then we determine $\sigma_1$ using
Eq.~\eqref{sigma_full}; (iii) After that we rescale the data  by
some factor $\mathcal{K}(\omega, T)$ determined in such a way
that the $\sigma_1
(H)$ maxima coincide with those obtained from either acoustic or DC
measurements.

To verify the suggested procedure, we applied it to both AS and MWS
for closely similar frequencies. The result is shown in
Fig.~\ref{fig:fig4_new}c. Since the frequencies are close, the
curves should coincide. This can indeed be seen to be the case, such
that our can be considered consistent.

\subsubsection{Frequency dependence of AC conductance}
\begin{figure}[h!]
\centering
\includegraphics[width=.9\columnwidth]{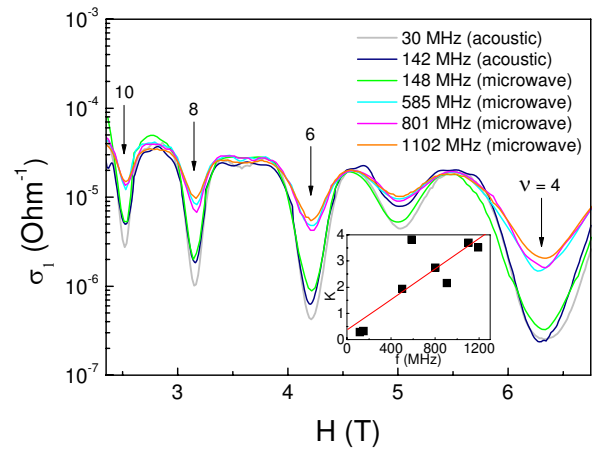}
\caption{Magnetic field dependences of $\sigma_1$ for different
frequencies at $T=1.7$~K.  The inset shows the frequency dependence of the scaling factor $\mathcal{K}$.
 \label{fig7_new}}
\end{figure}

\begin{figure}[ht]
\centering
\includegraphics[width=0.9\columnwidth]{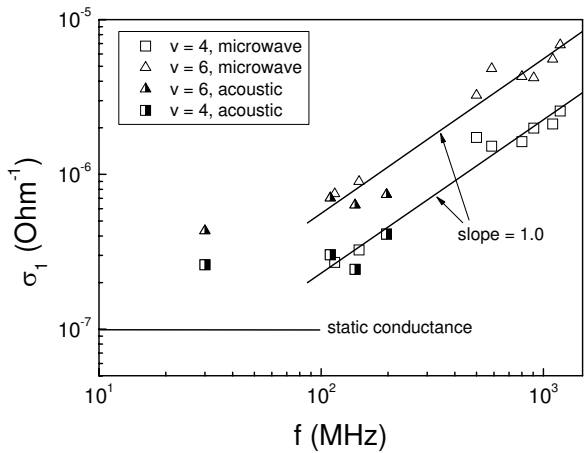}
\caption{Frequency dependence of the conductance in the minima
with $\nu =4$ and 6 at $T=1.7$~K.  The solid horizontal line shows the DC conductance, which is
practically the same for both minima.
  \label{fig8_new}}
\end{figure}

In Fig.~\ref{fig7_new} are shown the magnetic field dependences of
$\sigma_1$ for different frequencies. They are obtained using the
procedure outlined in Sec.~\ref{em}. The insert shows the frequency
dependence of
 the scaling factor $\mathcal{K}$.
%
The frequency dependence of the conductance in the minima with $\nu
=4, \, 6$ and at $T=1.7$~K is shown in
 Fig.~\ref{fig8_new}.
It is clear that at sufficiently high frequencies, $f \gtrsim
100$~MHz, the minimal values of $\sigma_1 (\omega)$ are roughly
proportional to  $\omega$, as it should be for AC hopping
conductance.  The solid line shows the value of the static
conductance, $\sigma_0$. At very low frequencies, the two-site model
leading to $\sigma_1 \propto \omega$ dependence is not applicable,
and more complicated clusters become important, see for a review
Refs.~\onlinecite{Efros1985,Dyre2000}.  As a result,
$\sigma_1(\omega)\to \sigma_0$.  This might be the reason of the
fact that the point corresponding  to $f=30$~MHz falls above the
line corresponding to the slope of 1 in the $\log \sigma_1$ vs.
$\log f$ dependence.

\section{Discussion and conclusions} \label{discussion}

We have developed a procedure for quantitative probeless measurements
of the AC conductance of 2D electron/hole layers in a broad frequency
domain.  The main ingredient of this procedure is to measure the
attenuation of surface acoustic waves (at low frequencies) and
electromagnetic modes in the CPW (at high frequencies)  in a
transverse magnetic field in the regime of the IQHE. Since the
transverse magnetic field  suppresses the electronic contribution to
the conductance both in diffusive and hopping regimes, it is
possible to resolve the contribution of the charge carriers.

Another important point is to rescale  the data obtained by MWS
such that the maxima of $\sigma_1(H)$ for all frequencies coincide
with the static conductance (and with the results of low-frequency
acoustic measurements). This rescaling to some extent compensates
for the leakage of the electromagnetic modes outside the slots of
the CPW.  The results of acoustic measurements do not need such
rescaling since the maxima of $\sigma_1(H)$ for all SAW frequencies
coincide with those of the static conductance. This fact allows
avoiding DC measurements which would require contacts. On the other
hand, $\sigma_1(H)$ extracted from MWS for any frequency can be
rescaled to make the maxima coinciding with those extracted from AS.
In this way, we can determine $\sigma_1(\omega,H)$ in a broad
domain of frequencies and magnetic fields without the need of
contacts. This is the main conclusion of this work.

The suggested procedure has been tested using a well-characterized sample.
It is shown that at frequencies close to 150~MHz, where both the AS
and MWS can be performed, the dependences $\sigma_1(H)$ obtained by
both spectroscopies practically coincide with each other.

The advantage of the procedure is that it can be applied to various
materials and structures. In particular, systems without intrinsic
piezoelectric effect can be studied acoustically
 since the sample is mounted on the surface of a piezoelectric crystal. The procedure is especially
 useful for studies of  the AC conductance in the hopping regime, which is the case in the minima of the IQHE.

\acknowledgments This work was supported by Russian Foundation for
Basic Research grant of RFBR 14-02-0023214, Presidium of the Russian
Academy of Science, the U.M.N.I.K grant 16906, and, partially, from
Era.Net-Rus.

\appendix
\section{Expressions for sound absorption and velocity} \label{A}
\begin{eqnarray}
&&   \frac{\Delta \Gamma\,
  (\text{dB/cm})}{8.68kA(k,a,d)}=\frac{\Sigma_1(B)}{[1+\Sigma_2(B)]^2+
  \Sigma_1^2(B)}
\nonumber \\ && \qquad \qquad \qquad \ \qquad
-\frac{\Sigma_1(0)}{[1+\Sigma_1(0)]^2+
  \Sigma_1^2(0)}\, ,  \label{eq:01} \\
&&\frac{v(B)-v(0)}{v(0)A(k,a,d)}=\frac{1+\Sigma_2(B)}{[1+\Sigma_2(B)]^2+
  \Sigma_1^2(B)}
\nonumber \\ && \qquad \qquad \qquad \qquad
-\frac{1+\Sigma_2(0)}{[1+\Sigma_1(0)]^2+
  \Sigma_1^2(0)}  \label{eq:02}
\end{eqnarray}
where
\begin{eqnarray}
&&A(k,a,d)= 110.2b(k,a,d)e^{-2k(a+d)}\, ,  \nonumber \\
&&\Sigma_i=4\pi
  t(a,k,d)\sigma_i/\varepsilon_sv(0)\, ; \nonumber \\
&&
b(k)=(b_1(k)[b_2(k)-b_3(k)])^{-1}
  \, , \nonumber \\
&&t(k,a,d)=[b_2(k)-b_3(k)]/2b_1(k)\, , \nonumber \\
&& b_1(k,a)=(\varepsilon_1+\varepsilon_0)(\varepsilon_s+\varepsilon_0)
\nonumber \\ && \quad \quad \quad
- (\varepsilon_1-\varepsilon_0)
(\varepsilon_s-\varepsilon_0)e^{-2ka}\, ,  \nonumber \\
&&
b_2(k,d)=(\varepsilon_1+\varepsilon_0)(\varepsilon_s+\varepsilon_0)
\nonumber \\ && \quad \quad \quad
+ (\varepsilon_1+\varepsilon_0)
(\varepsilon_s-\varepsilon_0)e^{-2kd}\, ,  \nonumber \\
&&
b_3(k,a,d)=
(\varepsilon_1-\varepsilon_0)(\varepsilon_s-\varepsilon_0)e^{-2ka}
\nonumber \\ && \quad \quad \quad
+(\varepsilon_1-\varepsilon_0)
(\varepsilon_s+\varepsilon_0)e^{-2k(a+d)}\, ,
\end{eqnarray}
$k$ is the SAW wave vector, $d$ is the depth of the 2D-system
layer in the sample, $a$ is the clearance between the sample and the
LiNbO$_3$ surface; $\varepsilon_1$=50, $\varepsilon_0$=1 and
$\varepsilon_s$=11.7 are the dielectric  constants of LiNbO$_3$, of
 vacuum, and of the semiconductor, respectively.

\end{document}